\documentclass[%
 reprint,
superscriptaddress,
 amsmath,amssymb,
 aps,
]{revtex4-2}

\usepackage{graphicx}% Include figure files
\usepackage{dcolumn}% Align table columns on decimal point
\usepackage{bm}% bold math
\usepackage[utf8]{inputenc}
\usepackage{graphicx}% Include Figure files
\usepackage{dcolumn}% Align table columns on decimal point
\usepackage{bm}% bold math
\usepackage{bbold}
\usepackage{physics}
\usepackage{mathtools} 
\usepackage{multirow}
\usepackage{hyperref}
\usepackage{makecell}
\usepackage{cleveref}

\usepackage{xcolor}

\usepackage[normalem]{ulem}

\begin{document}
\title{Diffusion disorder in the contact process}

\author{Valentin Anfray}
\email{valentin.anfray@gmail.com} % Optional
\affiliation{Institute of Physics, Academia Sinica, Taipei 115201, Taiwan}

\author{Manisha Dhayal}
\email{mdw6f@mst.edu}
\affiliation{Department of Physics, Missouri University of Science and Technology, Rolla, MO 65409, USA}

\author{Hong-Yan Shih}
\email{hongyan@as.edu.tw}
\affiliation{Institute of Physics, Academia Sinica, Taipei 115201, Taiwan}
\affiliation{Physics Division, National Center for Theoretical Sciences, Taipei 106319, Taiwan}

\author{Thomas Vojta}
\email{vojtat@mst.edu}
\affiliation{Department of Physics, Missouri University of Science and Technology, Rolla, MO 65409, USA}

\begin{abstract}
We study the effects of spatially inhomogeneous diffusion on the non-equilibrium phase transition in the contact process. The directed-percolation critical point in the contact process is known to be stable against the addition of a spatially uniform diffusion term. Correspondingly, we find quenched randomness in the diffusion rates to be irrelevant by power counting in the field-theory of the contact process. However, large-scale Monte Carlo simulations demonstrate that such diffusion disorder destabilizes the clean directed percolation critical point. Instead, the transition belongs to the same infinite-randomness universality class as the contact process with disorder in the infection or healing rates. To explain these results, we develop an effective model with an infinite diffusion rate; it shows that diffusion disorder generates an effective disorder in the healing rates. The same mechanism also appears in the field-theoretic description: Whereas diffusion disorder is irrelevant by power-counting, it generates standard random-mass disorder under renormalization. We discuss the validity of this mechanism for other absorbing state transitions and non-equilibrium phase transitions in general.
\end{abstract}

\maketitle

\section{Introduction}

Quenched disorder in transport properties has long been studied in statistical physics, particularly in the context of random walks in heterogeneous environments where spatial inhomogeneities are frozen on the timescale of the dynamics. Prominent examples include trap models, in which a particle escapes from site-dependent traps characterized by random waiting times $\mathcal{T}_i$. These systems exhibit normal diffusion when $\langle \mathcal{T}_i \rangle$ is finite and anomalous or ultraslow diffusion otherwise~\cite{bouchaud_anomalous_1990}. Related phenomena arise in random-force landscapes, such as the Sinai walk and its biased variants, where quenched heterogeneities lead to logarithmically slow dynamics or anomalous diffusion \cite{bouchaud_anomalous_1990, igloi_strong_2005}. 

The role of quenched disorder in diffusion has also been explored in interacting many-body systems far from equilibrium. A prominent example is the asymmetric exclusion process \cite{derrida_exactly_1998}, a prototypical driven diffusive model originally introduced to describe traffic flow \cite{nagel_cellular_1992, chou_non-equilibrium_2011}. Both particle-wise and site-wise disorder in the hopping rates have been studied \cite{evans_bose-einstein_1996, juhasz_partially_2006}, revealing collective effects such as the emergence of regimes with vanishing stationary current. 

In parallel, the influence of quenched disorder on critical phenomena has been extensively studied in both equilibrium and nonequilibrium settings, for reviews see, e.g., Refs.\ \cite{dotsenko_critical_1995, igloi_strong_2005, vojta_rare_regions_2006, *vojta_aip_proceedings_2013, *vojta_disorder_2019}. A central issue in this context is the relevance of disorder at a critical point. For weak, uncorrelated disorder coupled to the square of the order parameter (often referred to as random-mass disorder or  random-$T_c$ disorder), the stability of the clean critical behavior is governed by the Harris criterion, $d\nu_\perp>2$, where $\nu_\perp$ is the correlation-length exponent of the pure system and $d$ its spatial dimension \cite{harris_effect_1974}. Such disorder can arise, e.g., from random dilution of sites or bonds, or from spatial inhomogeneities of a control parameter. 

In contrast, quenched disorder in diffusion rates has received comparatively little attention. One possible reason for this neglect lies in a simple renormalization-group (RG) argument. In coarse-grained field-theoretic descriptions of reaction–diffusion systems, the scaling dimension of the product of the field and its Martin–Siggia–Rose response field generally satisfies  $[\phi\tilde{\phi}]=d$ \cite{tauber_critical_2014}. As a consequence, uncorrelated random spatial fluctuations of the diffusion coefficient acquire a negative scaling dimension under power counting and are therefore expected to be irrelevant. This reasoning suggests that quenched diffusion disorder should not affect the universal properties of non-equilibrium critical points.

Despite this expectation, diffusion is known to play a crucial role in several reaction–diffusion processes. In one-dimensional systems with nearest-neighbor infection and triplet annihilation, for instance, the absorbing phase disappears when the diffusion rate exceeds a critical value \cite{dickman_universality_1989}. In the diffusive epidemic process (DEP), the relative diffusion rates of the two species determine the universality class of the transition \cite{van_wijland_wilson_1998}. Recent work has further shown that quenched heterogeneities in the diffusion rates of DEP change the critical properties and can even lead to new phenomena that are absent when disorder is introduced solely through the control parameter \cite{anfray_relevance_2025}. These results challenge the naive irrelevance argument and raise questions about the effect of diffusion disorder in nonequilibrium systems at criticality.

Motivated by this apparent contradiction, we revisit the effect of diffusion on the contact process (CP) \cite{harris_contact_1974}, a paradigmatic nonequilibrium model exhibiting an absorbing-state phase transition \cite{hinrichsen_nonequilibrium_2000, odor_universality_2004, henkel_non-equilibrium_2008} in the directed percolation (DP) \cite{grassberger_directed_1989} universality class. Specifically, our work examines whether quenched spatial disorder in the diffusion rates can modify the critical behavior. After introducing the CP with quenched diffusion disorder in Sec.\ \ref{sec:model}, we first analyze the relevance of diffusion disorder using power counting. In Sec.\ \ref{sec:MC}, we present extensive Monte Carlo simulations of the CP with diffusion disorder. To elucidate the underlying physics, we derive an effective description showing how diffusion disorder couples to other reaction rates in Sec.\ \ref{sec:effective}. Section \ref{sec:Feynman} is devoted to the effects of disorder renormalizations by higher-order Feynman diagrams in the field-theoretic description. Our conclusions are summarized in the final section.

%%%%%%%%%%%%%%%%%%%%%%%%%%%%%%%%%%%%%%%%%%%%%%%%%%%%%%%%%%%%%%%%
\section{Model and field theory}
\label{sec:model}
\subsection{Contact Process}

Throughout this paper, we consider the contact process (CP) \cite{harris_contact_1974} to study the effect of quenched diffusion disorder on the directed percolation (DP) universality class. The CP can be understood as a minimal stochastic lattice model for the spreading of an infection.  It is defined on a $d$-dimensional hypercubic lattice where each site can be either active (infected) or inactive (healthy). Active sites spontaneously become inactive at rate $\tau$ and infect each of their $2d$ nearest neighbors at rate $\lambda/(2d)$. The CP exhibits a continuous phase transition between an absorbing (inactive) phase devoid of active sites and an active phase characterized by a nonzero steady-state density $\rho$ of active sites. The transition is controlled by the ratio of the infection and healing rates, $\lambda/\tau$. Its critical behavior is known to belong to the DP universality class.

To study diffusion, we add a third basic process to the CP: We explicitly allow the infection to hop from an active site to one of its neighbors at rate $D$. Such a process is already implicitly contained in the usual CP through an infection event, for example when an active site $i$ infects a neighbor $j$, followed immediately by a healing event of site $i$. It has been verified that adding such explicit diffusion is an irrelevant perturbation at the critical point of the CP and therefore does not modify the critical behavior \cite{jensen_time-dependent_1993}.

We now introduce uncorrelated quenched spatial diffusion disorder. Each lattice site $i$ is assigned a diffusion rate $D_i$ drawn from some probability density, and the $D_i$ at different sites are independent of each other.

\subsection{Power counting}
To predict whether quenched diffusion disorder is a relevant perturbation, it is convenient to examine the field-theoretic functional of the CP. Since the critical behavior of the CP belongs to the DP universality class, we start from its dynamic functional \cite{janssen_renormalized_1997,hinrichsen_nonequilibrium_2000,henkel_non-equilibrium_2004}
\begin{equation}
    S[\tilde{\phi},\phi] = \int d^dxdt \tilde{\phi}\left(\partial_t - D\nabla^2 - r + \frac{\Gamma}{2}(\phi - \tilde{\phi})\right)\phi ,
    \label{eq:DPaction}
\end{equation}
where $\phi(\boldsymbol{x},t)$ is the density field of active sites at position $\boldsymbol{x}$ and time $t$, $\tilde{\phi}(\boldsymbol{x},t)$ is the Martin–Siggia–Rose response field~\cite{martin_statistical_1973},  $\nabla^2$ is the Laplacian, $r$ controls the deviation from criticality, and $\Gamma$ is a nonlinear coupling constant. For the CP, these parameters are related to $\lambda$ and $\tau$ via $r=\lambda-\tau$ and $\Gamma=\lambda$. To include quenched disorder in the diffusion coefficient, we write $D(\boldsymbol{x}) = D_0 + \chi_{D}(\boldsymbol{x})$, where $D_0$ is a constant and $\chi_{D}(\boldsymbol{x})$ is a Gaussian distributed random variable with mean and covariance given by
\begin{equation}
    \langle \chi_D(\boldsymbol{x}) \rangle =0, \quad \langle \chi_D(\boldsymbol{x})\chi_D(\boldsymbol{x'})\rangle = \frac{\sigma_D}{2} \delta(\boldsymbol{x}-\boldsymbol{x'}).
\end{equation}
In contrast, randomness in the infection and/or healing rates leads to random-mass disorder which can be included by writing $r(\boldsymbol{x}) = r_0 + \chi_r(\boldsymbol{x})$ \cite{janssen_renormalized_1997}. Here, $r_0$ is a constant and $\chi_r(\boldsymbol{x})$ is a Gaussian distributed random variable with zero mean and covariance $\sigma_r/2$.

Substituting \(D(\boldsymbol{x})\) and \(r(\boldsymbol{x})\) into Eq.~\eqref{eq:DPaction}, one obtains
\begin{align*}
    S[\tilde{\phi},\phi] =& \int d^dxdt \tilde{\phi}\left(\partial_t - D_0 \nabla^2 - r_0 \right)\phi \\
    &+ \int d^dxdt \frac{\Gamma}{2}\tilde{\phi}(\phi-\tilde{\phi}) \phi \\
    &- \int d^dxdt\chi_D(\boldsymbol{x}) \tilde{\phi}\nabla^2 \phi \\
    &- \int d^dxdt\chi_r(\boldsymbol{x}) \tilde{\phi} \phi \\
    =& S_0 + S_{int}+S_{\text{dis},\chi_D} +S_{\text{dis},\chi_r} .
\end{align*}
Averaging $\exp(-S)$ over the quenched disorder can be done directly without the need to use the replica trick \cite{janssen_renormalized_1997} and yields the replacement of $S_{\text{dis},\chi_D}$ and $S_{\text{dis},\chi_r}$ with 
\begin{align}
    S_{\text{dis},D} &= \int d^dx  \frac{\sigma_D}{2} \left[\int dt \tilde{\phi}\nabla^2 \phi\right]^2, \label{eq:Sdis_D} \\
    S_{\text{dis},r} &= \int d^dx  \frac{\sigma_r}{2} \left[\int dt \tilde{\phi} \phi\right]^2 . \label{eq:Sdis_r}
\end{align}

We now analyze the relevance of $S_{\text{dis},D}$ and $S_{\text{dis},r}$ by power counting. Applying the usual rescaling of lengths $x \to bx$ and times $t \to D_0b^{z} t$ (with the Gaussian fixed-point value $z=2$), the scaling dimensions are
\begin{equation}
    [\phi] = [\tilde{\phi}] = \frac{d}{2}, \quad [\Gamma] = \frac{1}{2}(4-d), \quad [\sigma_D] = -d, \quad [\sigma_r] = 4-d .
\end{equation}
The equality $[\phi]=[\tilde{\phi}]$ follows from the rapidity-reversal symmetry \cite{hinrichsen_nonequilibrium_2000,henkel_non-equilibrium_2004}. Power counting shows that $\Gamma$ is relevant for $d<4$, identifying $d_c=4$ as the upper-critical dimension of the CP in the absence of quenched disorder. In contrast, disorder in the diffusion coefficient always has a negative scaling dimension and is therefore irrelevant for all $d$.
This situation differs from the case of random-mass disorder which is relevant for $d<4$. A perturbative RG study of the field theory with random-mass disorder did not produce a stable fixed point \cite{janssen_renormalized_1997}. Instead, a strong-disorder renormalization group (SDRG) calculation \cite{hooyberghs_strong_2003,hooyberghs_absorbing_2004} predicted an exotic infinite-randomness fixed point 
(IRFP) in the same universality class as the random transverse-field Ising model \cite{fisher_critical_1995,MMHF_2000,kovacs_renormalization_2010,kovacs_infinite-disorder_2011}.  This has been confirmed by large scale Monte Carlo simulations of the CP with random-mass disorder in one, two, and three dimensions \cite{vojta_critical_2005,vojta_infinite-randomness_2009,vojta_monte_2012}.

Thus, power counting suggests that quenched heterogeneities in the diffusion coefficient are irrelevant, in contrast to heterogeneities in the mass term.  Applied to the CP, this implies that randomness in the hopping rates $D_i$ should not affect the critical behavior. However, because a hopping event can be viewed as an infection event immediately followed by a healing event, one may also expect that diffusion disorder contributes to a spatial modulation of the effective local infection and/or healing rates, which would correspond to a relevant perturbation. This apparent contradiction motivates the detailed investigation presented below.

%%%%%%%%%%%%%%%%%%%%%%%%%%%%%%%%%%%%%%%%%%%%%%%%%%%%%%%%%%%%%%%%%%
\section{Monte Carlo Simulations}
\label{sec:MC}
\subsection{Simulation method}
We have performed Monte Carlo simulations of the one-dimensional CP with quenched diffusion disorder. The disorder is implemented by making the local diffusion rate $D_{i}$ of lattice site $i$ an independent random variable. The probability density of the local diffusion rate is binary and is given by
\begin{equation}
P(D_i)=(1-p)\delta(D_i-D)+ p\delta(D_i).
\end{equation}
Here, $p$ denotes the spatial density of sites with zero local diffusion rate, and $1-p$ is the density of sites with diffusion rate $D$. The simulations model the spreading of an infection and begin with infecting a single randomly chosen site from an empty (healthy) lattice at time $\textit{t}=0$. 

The simulations are performed using the algorithm by Dickman~\cite{dickman_reweighting_1999} and consist of a sequence of events. Each event starts by randomly selecting a site from a list of all $N_{s}$ infected (occupied) sites, after which we choose a process at random from one of the following: infection with probability $\lambda/(\tau+\lambda +\textit{D})$, healing with probability $\tau/(\tau+\lambda +\textit{D})$, and diffusion with probability $\textit{D}/(\tau+\lambda +\textit{D})$. 
In the case of infection, one of the neighboring sites is picked randomly and is infected if empty. In the case of diffusion, the infection hops to one of the neighboring sites, chosen randomly, if it is empty and if the current site has diffusion rate $D_{i}=D$; otherwise, the infection stays at the current site. The process of healing turns the infected site into a healthy one. The time increment of this event is given by $1/N_{s}$.

Using this algorithm, we have run simulations for time periods ranging from $10^{6}$ to $3\times10^{8}$. The data are averaged over a large number of disorder realizations, ranging from $10^{5}$ for the shorter times to 24600 for longest. For each realization we make $20$ attempts to grow the infection. In the production runs, we have chosen the diffusion rate $\textit{D}=10$ with the impurity concentration $p=0.25$ and the healing rate $\tau=1$. The infection rate $\lambda$ is used to tune the phase transition. These parameters are chosen to implement strong disorder. We will return to this question in Sec.\ \ref{sec:MCresults}. The system sizes vary from $L=2\times10^{6}$ to $2\times10^{8}$. For these values, the active cluster remains well within the system boundaries at all times, avoiding any finite size effects. 

\subsection{Scaling Theory}
In this section, we provide a brief overview of the scaling theories for nonequilibrium phase transitions in the clean CP and in the CP with disorder in the infection or healing rates. These scaling theories will be used to analyze our Monte Carlo simulation results and determine the critical behavior. Further details can be found in Ref.~\cite{hinrichsen_nonequilibrium_2000} for power-law scaling and in Ref.~\cite{vojta_critical_2005} for activated scaling.
The central quantity in the CP is the average density of active sites at time $t$
\begin{equation}
\rho(t) = \frac{1}{L}\sum_{i}\langle n_{i}(t)\rangle
\end{equation}
where $n_{i}(t)$ is the site occupation function which is $1$ if the site $i$ is infected at time $t$ and zero otherwise. $L$ is the linear system size, and $\langle...\rangle$ denotes the average over all Markov processes.
In absorbing state transitions, the order parameter is given by the steady state density
 \begin{equation}
\rho_{st}=\lim_{t \to \infty}\rho(t) ~.
\end{equation}
Near the critical infection rate $\lambda_{c}$ in the active phase, the steady state density varies as
\begin{equation}
\rho_{st}\sim (\lambda -\lambda_{c})^{\beta}\sim r^{\beta} 
\end{equation}
where $r=(\lambda-\lambda_{c})/\lambda_{c}$ is the dimensionless distance from the critical point and $\beta$ is the critical exponent associated with the particle density.

\subsubsection{Conventional critical points} 

The dynamics of the clean CP near criticality is controlled by a conventional critical point~\cite{hinrichsen_nonequilibrium_2000} and therefore follows power law scaling. The characteristic correlation length $\xi_{\perp}$ diverges at the critical point as 
\begin{equation}
\xi_{\perp}\sim \lvert r\rvert^{-\nu_{\perp}} 
\label{eq:correlation_length}
\end{equation}
and the correlation time $\xi_{\parallel}$ diverges as a power of $\xi_{\perp}$,
\begin{equation}
\xi_{\parallel}\sim \xi_{\perp}^{z} 
\end{equation}
where $z$ is the dynamical exponent. The scaling forms of the density and various quantities measured in spreading runs (starting from a single infected site) as functions of the critical distance $r$, system size $L$ and time $t$, can be expressed in terms of $\beta$, $\nu_{\perp}$ and $z$. The particle density scales as  
\begin{equation}
    \rho(r,t,L)= b^{\beta/\nu_{\perp}}\rho(r b^{-1/\nu_{\perp}},tb^{z},Lb)                            
\end{equation}
where $b$ is an arbitrary length scale factor. The survival probability $P_{s}$ has the same scaling form as the particle density (this follows from the rapidity reversal symmetry),
\begin{equation}
P_{s}(r,t,L)= b^{\beta/\nu_{\perp}}P_{s}(r b^{-1/\nu_{\perp}},tb^{z},Lb).
\end{equation}
The mean square radius of the cluster of infected sites (when starting from a single seed site) scales as a length; it therefore obeys the relation 
\begin{equation}
R(r,t,L)= b^{-1}R(r b^{-1/\nu_{\perp}},tb^{z},Lb).
\end{equation}
The correlation function $\tilde C(i_{1},t_{1},i_{2},t_{2})=\langle n_{i_{1}}(t_{1})n_{i_{2}}(t_{2})\rangle$ is defined as the probability that site $i_{1}$ is active at time $t_{1}$ if site $i_{2}$ is active at time $t_{2}$. Since the correlation function is translationally invariant in space and time, it depends only on the distance $i=i_1-i_2$ between the sites and the time $t=t_1-t_2$ between the two events, $\tilde C(i_{1},t_{1},i_{2},t_{2})=C(i,t)$. Since $C$ involves a product of two densities, it scales as~\cite{hyperscaling_note}
\begin{equation}
C(r,i,t,L)= b^{2\beta/\nu_{\perp}}C(r b^{-1/\nu_{\perp}}, ib,tb^{z},Lb).
\end{equation}
The total number of active sites $N_{s}$ in the surviving cluster (when starting from a single seed site) is obtained by integrating the correlation function over all space, giving 
\begin{equation}
N_{s}(r,t,L)= b^{2\beta/\nu_{\perp}-d}N_{s}(r b^{-1/\nu_{\perp}},tb^{z},Lb)
\end{equation}
where $d=1$ is the space dimensionality.
In the long-time limit at the critical point, $r=0$, and in the thermodynamic limit $L\to \infty$, the observables follow the power laws given below:
\begin{align}
\rho(t)\sim t^{-\delta},&\quad P_{s}(t)\sim t^{-\delta},\\
R(t)\sim t^{1/z},&\quad N_{s}\sim t^{\Theta}
\end{align}
where $\delta=\beta/(\nu_{\perp}z)$, and $\Theta=d/z-2\beta/(\nu_{\perp}z)$ is the so-called critical initial
slip exponent. The critical exponents of the clean one-dimensional CP (DP universality class) are known to high precision from series expansion calculations ~\cite{jensen_low-density_1999} as follows: $\beta = 0.276486$, $\nu_{\perp} = 1.096854$, $z = 1.580745$, $\delta = 0.159464$, $\Theta = 0.313686$.

\subsubsection{Infinite-randomness critical points} 

In the case of the CP with quenched disorder in the infection rates, the phase transition was shown to be controlled by an IRFP featuring activated scaling~\cite{hooyberghs_strong_2003,hooyberghs_absorbing_2004,vojta_critical_2005,vojta_infinite-randomness_2009,vojta_monte_2012}. Infinite-randomness critical points exhibit extremely slow dynamics, i.e., the correlation time $\xi_{\parallel}$ depends exponentially on the correlation length $\xi_{\perp}$ as
\begin{equation}
  \ln(\xi_{\parallel}/t_{0})\sim\xi_{\perp}^{\psi}
  \label{eq:activated_correlation_time}
\end{equation}
where $\psi$ is the so-called tunneling exponent and $t_{0}$ is a nonuniversal microscopic time scale.
The scaling relations for the observables are modified by replacing the time dependence $tb^{z}$ with $\ln(t/t_{0})b^{\psi}$  as follows
\begin{align}
  \rho(r,\ln(t/t_{0}),L)&= b^{\beta/\nu_{\perp}}\rho(r b^{-1/\nu_{\perp}},\ln(t/t_{0})b^{\psi},Lb))\label{eq:activated_scaled_rho}\\
  P_{s}(r,\ln(t/t_{0}),L)&= b^{\beta/\nu_{\perp}}P_{s}(r b^{-1/\nu_{\perp}},\ln(t/t_{0})b^{\psi},Lb))\label{eq:activated_scaled_Ps}\\
  R(r,\ln(t/t_{0}),L)&= b^{-1}R(r b^{-1/\nu_{\perp}},\ln(t/t_{0})b^{\psi},Lb))\label{eq:activated_scaled_R}\\
  N_{s}(r,\ln(t/t_{0}),L)&= b^{2\beta/\nu_{\perp}-d}N_{s}(r b^{-1/\nu_{\perp}},\ln(t/t_{0})b^{\psi},Lb)\label{eq:activated_scaled_Ns}.
\end{align}
 The time dependencies of the observables at criticality in the thermodynamic limit are now logarithmic, 
\begin{align}
\rho(t)\sim \ln(t/t_{0})^{-\bar{\delta}},&\quad P_{s}(t)\sim \ln(t/t_{0})^{-\bar{\delta}},   \label{eq:asymptotic_rho_Ps}\\
R(t)\sim \ln(t/t_{0})^{1/\psi},&\quad N_{s}\sim \ln(t/t_{0})^{\bar{\Theta}}   \label{eq:asymptotic_R_Ns},
\end{align}
where $\bar{\delta}=\beta/(\nu_{\perp}\psi)$ and $\bar{\Theta}=d/\psi-2\beta/(\nu_{\perp}\psi)$. 
Exact values of the critical exponents for the disordered one-dimensional CP have been calculated by the SDRG approach ~\cite{hooyberghs_strong_2003,hooyberghs_absorbing_2004}. They read $\beta=(3-\sqrt{5})/2=0.38197$, $\nu_{\perp}=2$, $\psi=0.5$, $\bar{\delta}=0.38197$, and $\bar{\Theta}=1.2360$.

\subsection{Results}
\label{sec:MCresults}

To test our numerical implementation of the CP, we first perform short test simulations of the clean CP in the absence of diffusion. We find a critical infection rate of $\lambda_c \approx 3.30$, in agreement with the literature \cite{vojta_critical_2005}. We also study
a system with a uniform diffusion rate of $D=10$. Its critical infection rate, $\lambda_c \approx 2.04$, is much lower, but the critical behavior of the phase transition agrees with the clean DP universality class, as expected  \cite{jensen_time-dependent_1993}.

We now turn to the simulations with diffusion disorder. In order to make the effects of diffusion disorder clearly visible, we aim at implementing strong disorder. A large value, $D=10$, of the diffusion rate creates a significant difference between sites having nonzero diffusion, $D_i=D$, and sites with $D_i=0$. The above test calculations for the clean CP have shown that a nonzero $D$ shifts $\lambda_c$ downward. Choosing $D=10$ together with a relatively small concentration, $p=0.25$, of sites with zero diffusion rate thus allows for the formation of large active rare regions in an inactive bulk.

\begin{figure}
    \centering
    \includegraphics[width=\linewidth]{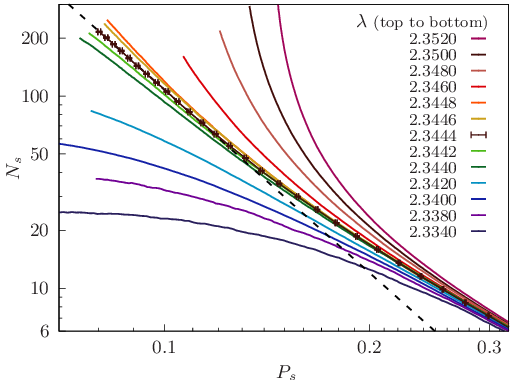}
    \caption{Number of active sites $N_{s}$ vs. survival probability $P_{s}$ for several infection rates $\lambda$. Data for the critical curve $\lambda_{c}=2.3444$ (shown with errorbars for select points) is fitted with Eq.\eqref{eq:activated_Ns_Ps} with a floating exponent,  yielding $\bar{\Theta}/\bar{\delta}=3.24(1)$. The fit is of good quality $(\chi^{2}\approx0.05)$ and shown with straight dashed line.}
   \label{fig:Ns_vs_Ps}
\end{figure}

To identify the critical point in the Monte Carlo results, we check the relation between the observables $N_{s}$ and $P_{s}$. This has the advantage that the relation between $N_{s}$ and $P_{s}$ is of power-law type for both conventional and activated scaling. For the clean one-dimensional DP critical point, we have
\begin{equation}
N_{s}\sim P_{s}^{-\Theta/\delta}
\label{eq:conventional_Ns_Ps}
\end{equation}
with the exponent $\Theta/\delta\approx1.9673$. For the infinite-randomness critical point of the CP with random-mass disorder, we obtain the relation 
\begin{equation}
N_{s}\sim P_{s}^{-\bar{\Theta}/\bar{\delta}}
\label{eq:activated_Ns_Ps}
\end{equation}
with the exponent $\bar{\Theta}/\bar{\delta}\approx3.2359$. Looking for a power law in $N_{s}$ vs.\ $P_{s}$ thus does not imply assumptions about the scaling scenario. In Fig.~\ref{fig:Ns_vs_Ps} we have graphed $N_{s}$ vs. $P_{s}$ on a log-log scale. 
We observe that the data for both $\lambda=2.3444$ and $2.3446$ are approximately linear for long times. The precise critical value likely lies between these two values. However, from our analysis we find $\lambda=2.3444$ to be a better approximation of the critical value because the data follows power-law behavior over a wider time range. From fitting the data for $\lambda=2.3444$ with Eq.\eqref{eq:activated_Ns_Ps}, we find the value of the exponent to be $\bar\Theta/\bar\delta=3.24(1)$, in agreement with the SDRG prediction~\cite{hooyberghs_strong_2003,hooyberghs_absorbing_2004} for the CP with random-mass disorder.  

\begin{figure}
    \centering
    \includegraphics[width=\linewidth]{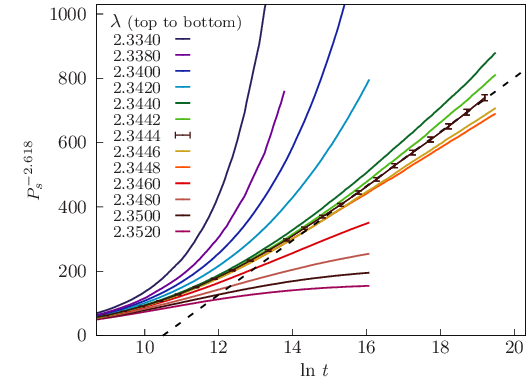}
    \caption{Survival probability $P_{s}^{-1/\bar\delta}$ vs. $\ln t$  for several infection rates $\lambda$. Data for the critical curve $\lambda_{c}=2.3444$ (shown with errorbars for select points) is fitted with Eq.\eqref{eq:asymptotic_rho_Ps} with fixed exponent ${1/\bar{\delta}}=2.618$ following. The fit is shown here with straight dashed line.}
   \label{fig:Asymptotic_Ps}
\end{figure}

\begin{figure}
    \centering
    \includegraphics[width=\linewidth]{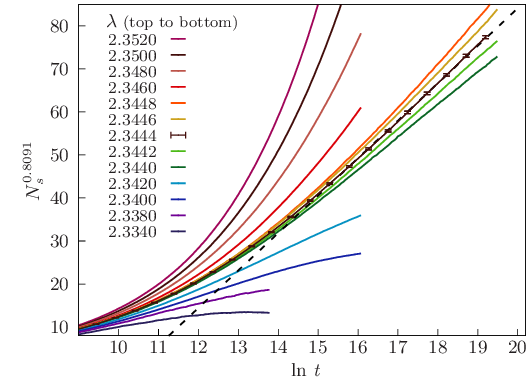}
    \caption{Number of active sites $N_{s}^{1/\bar\Theta}$ vs. $\ln t$  for several infection rates $\lambda$. Data for the critical curve $\lambda_{c}=2.3444$ (shown with errorbars for select points) is fitted with Eq.\eqref{eq:asymptotic_R_Ns}with fixed exponent $1/\bar\Theta=0.8091$. The fit is shown here with
    straight dashed line.}
   \label{fig:Asymptotic_Ns}
\end{figure}

We now check the time dependencies of $P_{s}$ and $N_{s}$ to confirm that the critical point is characterized by activated scaling. We graph $P_{s}^{-1/\bar{\delta}}$ vs. $\ln t$ in Fig.~\ref{fig:Asymptotic_Ps}, fixing ${1/\bar{\delta}}$ at the predicted value $1/\bar\delta=2.618$. We also plot $N_{s}^{1/\bar\Theta}$ vs. $\ln t$ with the predicted $1/\bar\Theta=0.8091$ in Fig.~\ref{fig:Asymptotic_Ns}. 
In these plots, Eqs.\eqref{eq:asymptotic_rho_Ps} and \eqref{eq:asymptotic_R_Ns} result in straight lines independent of the value of the microscopic timescale $t_{0}$. Fits of the critical curves ($\lambda=2.3444$) with $\bar\Theta$ and $\bar\delta$ fixed at the predicted values agree with the data for about two orders of magnitude in time. We also find estimates of the microscopic timescale $t_{0}$ from the fits. Fig.~\ref{fig:Asymptotic_Ps} gives $t_{0}\approx 10.49$, and Fig.~\ref{fig:Asymptotic_Ns} gives $t_{0}\approx10.34$. The close agreement between these independently determined values of $t_0$ at $\lambda_{c}=2.3444$, in contrast to a larger discrepancy observed at $\lambda=2.3446$, serves as an additional consistency check on the critical infection rate. The relation $\bar\Theta=d/\psi-2\bar\delta$ implies $\psi=0.5$ as is obtained in ~\cite{hooyberghs_absorbing_2004}. 

\begin{figure}
    \centering
    \includegraphics[width=\linewidth]{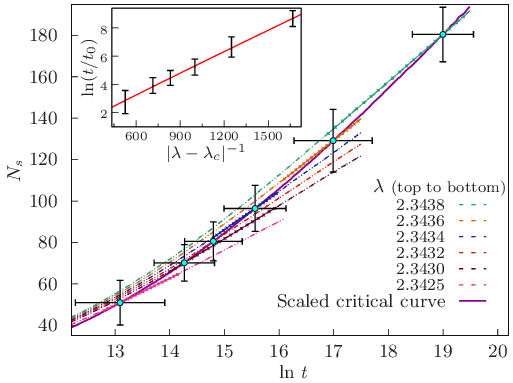}
    \caption{Number of active sites $N_{s}$ vs. $\ln t$ for several infection rates $\lambda$ in the inactive phase and scaled critical curve $0.86\times N_{s}$ at $\lambda=\lambda_{c}$. Linear fits have been used to find the intersection points (shown here with error bars) between the scaled critical curve and the subcritical curves.  Inset: Crossing time vs. $\lvert\lambda-\lambda_{c}\rvert^{-1}$. The solid line is a power-law fit, yielding $\nu_{\perp}\psi=0.96(2)$.}
   \label{fig:nu_val_Ntf}
\end{figure}

To accurately identify the remaining exponents $\beta$ and $\nu_{\perp}$ characterizing the universality class, we plot $N_{s}$ vs. $\ln t$ and scale the critical curve by a constant factor of 0.86. We then find the intersection points between the scaled critical curve and several curves (slightly) in the inactive phase. According to Eq.\eqref{eq:activated_scaled_Ns}, the intersection points are expected to follow $\ln(t/t_{0})\sim\lvert\lambda-\lambda_{c}\rvert^{-\nu_{\perp}\psi}$. This analysis is shown in Fig.~\ref{fig:nu_val_Ntf}. 
From the fit in the inset of Fig.~\ref{fig:nu_val_Ntf} we find the exponent $\nu_{\perp}\psi=0.96(2)$, in agreement with the SDRG prediction for the CP with random-mass disorder.

To confirm the universality of the critical behavior of the CP with quenched diffusion disorder, we perform additional simulations with hopping distance $3$. In this case, particles diffuse by hopping to the third-nearest site rather than the nearest neighbor, leading to a larger effective diffusion constant. These simulations produce the exact same critical behavior as described above.

In summary, our Monte Carlo simulations show that diffusion disorder is a relevant perturbation and destabilizes the critical point of the clean CP. Moreover, our Monte Carlo simulations also show that the CP with diffusion disorder belongs to the same universality class as the CP with quenched disorder in the infection rate~\cite{vojta_critical_2005}.    

%%%%%%%%%%%%%%%%%%%%%%%%%%%%%%%%%%%%%%%%%%%%%%%%%%%%%%%%%%%%%%%%%%%%%%%%%%%%%%%%%%%
\section{Effective model with infinite diffusion rate}
\label{sec:effective}
\subsection{Effective model}

To understand why quenched disorder in the diffusion rate can modify the critical behavior, we consider a binary distribution in which the local diffusion coefficient is either $0$ with probability $p$ or formally infinite. By "infinite" diffusion, we simply mean that the probability of finding an active site anywhere inside a region with $D=\infty$ is uniform. Because quenched disorder creates regions of finite size with the same diffusion value, whose probability decays exponentially with their size, one may relax the notion of "infinite" diffusion and instead take $D$ to be sufficiently large compared to the largest typical region. We emphasize that the purpose of this construction is not to propose a realistic microscopic model, but to obtain a more tractable model to better understand the effect of random diffusion.

\begin{figure}
    \centering
    \includegraphics[width=\linewidth]{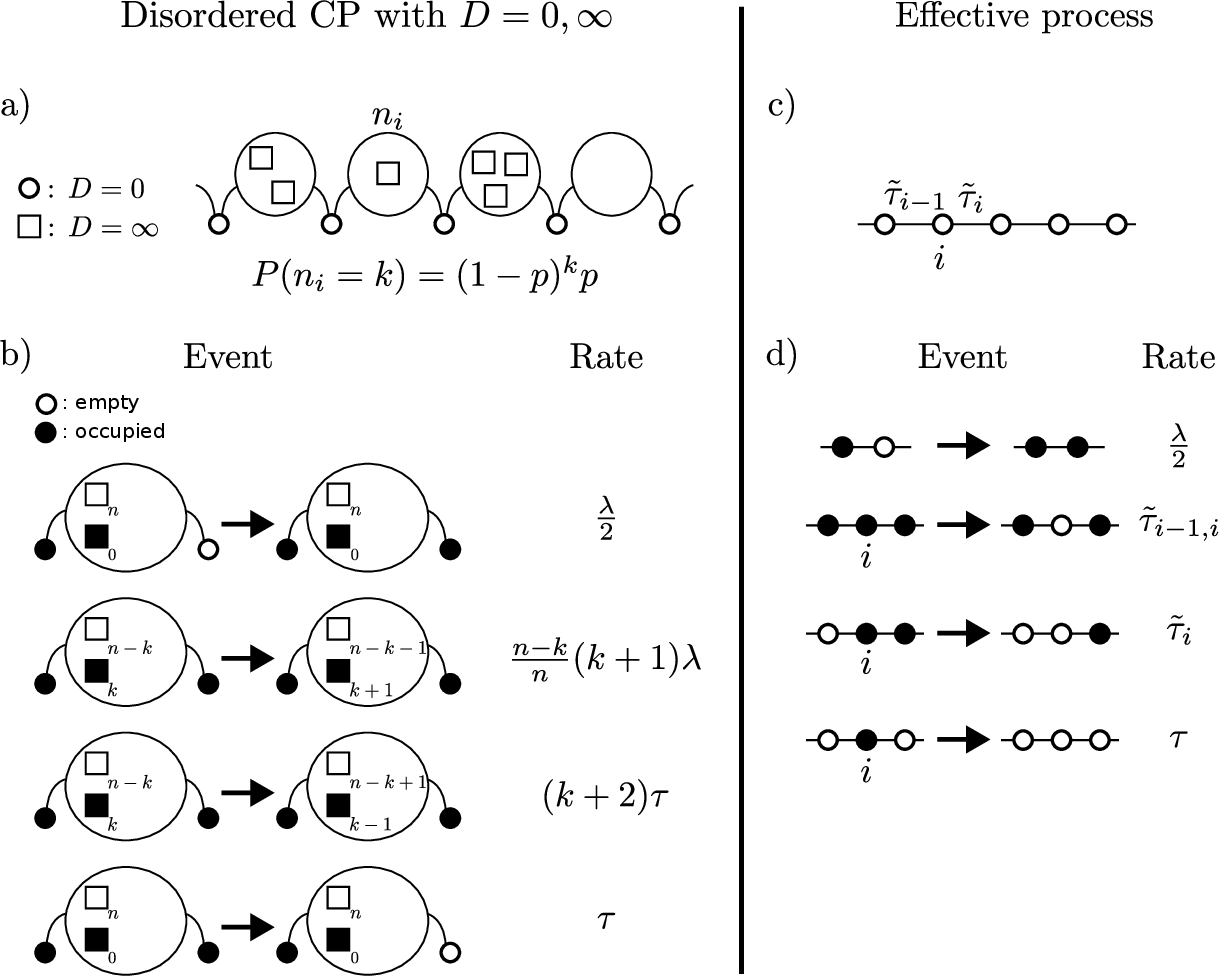}
    \caption{a) Contact process (CP) with quenched binary diffusion disorder: $D_i=0$ with probability $p$ and $D_i=\infty$ otherwise, where adjacent $D_i=\infty$ sites form a pocket. b) Dynamical rules. c) Effective dynamics of the $D_i=0$ sites, resembling a disordered CP with site-dependent healing rates. d) Dynamical rules of the effective process; when more than one neighbor is active, the survival probability is non-Poissonian and depends on the number of $D_i=\infty$ sites in between. The effective rates $\tilde{\tau}_i$ and $\tilde{\tau}_{i-1,i}$ denote the late-time exponential decay rates of the survival probability.}
    \label{fig:illustration_DinfCP}
\end{figure}
Within this binary disorder, the key observation is that an active site $i$ with $D_i=0$ effectively never becomes inactive as long as at least one of its neighboring regions defined by successive sites with $D=\infty$, called a pocket, contains at least one active site. This is because whenever a healing event occurs at a site $i$ with $D_i=0$, an active site in the adjacent pocket immediately hops to site $i$ and reactivates it. Motivated by this mechanism, we introduce the following effective model.

We consider a one-dimensional chain of sites that can be active or inactive. Between any two sites, we insert a pocket containing $n$ sites, where $n$ follows a geometric distribution, \(P(n=u) = (1-p)^u p\) with \(u\geq0\), see Fig.~\ref{fig:illustration_DinfCP}a). Each site in the pocket can also be active or inactive. Because diffusion inside a pocket is infinitely fast, it is sufficient to track only the number of active sites $k$ in the pocket, not their positions. We obtain the dynamical rules illustrated in Fig.~\ref{fig:illustration_DinfCP}b). All the rules follow straightforwardly from those of the CP, except for the computation of the infection rate inside a pocket, which we now detail.

Each pocket is surrounded by two $D_i=0$ sites, which we call external sites. An external active site infects a neighboring pocket at rate $\lambda/2$. The probability of success, i.e., that the pocket site adjacent to the external site is empty, is $(n-k)/n$. An active site inside the pocket is adjacent to an external active site with probability $2/n$ and then infects the opposite neighbor at rate $\lambda/2$, which is inactive with probability $1-(k-1)/(n-1)$. Otherwise, it attempts to infect a pocket site at rate $\lambda$, which is inactive with probability $(n-k)/(n-1)$. Summing all contributions gives
\[
2\frac{\lambda}{2}\frac{n-k}{n} + k\!\left( \frac{\lambda}{2}\frac{2}{n}\frac{n-k}{n-1} + \lambda\frac{n-2}{n}\frac{n-k}{n-1} \right)
= \lambda (k+1)\frac{n-k}{n}.
\]

Focusing solely on the $D_i=0$ sites, their dynamics resembles the contact process: a site can infect its neighbors or undergo a healing event. The difference lies in the healing event, which now depends on the size of the adjacent pockets and on whether the neighboring external sites are active or inactive, as illustrated in Fig.~\ref{fig:illustration_DinfCP}c)-d). If both neighbors are inactive, the site heals as in the CP with rate $\tau$. If site $i+1$ is active, the decay of site $i$ is slower; we denote the corresponding effective rate symbolically by $\tilde{\tau}_i$. However, $\tilde{\tau}_i$ is not a true rate since the survival probability is no longer exponential. For example, in Appendix~\ref{app:S_proba} we derive the exact survival probability for site $i$ with a neighboring pocket of size $n=1$
\begin{equation}
    S(t)=e^{-(\lambda+4\tau)t/2}\left(\cosh\!\frac{\Delta t}{2}+\frac{\lambda+2\tau}{\Delta}\sinh\!\frac{\Delta t}{2}\right).
\end{equation}
Only at late times does $S(t)$ decay exponentially. More generally, because pockets have finite size, the survival probability always decays exponentially at late times; we denote this decay rate by $\tilde{\tau}_i$ (see also Appendix~\ref{app:LiouvilleDecay}). Finally, if a site $i$ has two active neighbors, $S(t)$ still decays exponentially at late times, at a rate denoted by $\tilde{\tau}_{i-1,i}$ which is even smaller than both $\tilde{\tau}_i$ and $\tilde{\tau}_{i-1}$ due to the combined stabilizing effect of the two neighboring pockets. 

As a result, this effective model cannot be mapped exactly onto the CP with quenched disorder in the healing rates. However, as the healing events remain short-range and decay sufficiently fast (exponentially at late times), the critical properties are still expected to be identical to those obtained in the presence of random quenched healing rates.

\subsection{Simulation method}

To simulate the dynamical rules described in the previous section, we use the following algorithm. Time is measured in units of $1/(\lambda+\tau)$; it is initialized at $t=0$ and incremented by $1/N_s$ after each event, where $N_s$ is the number of active sites. Each event consists of randomly selecting one of the $N_s$ active sites and determining the type of event: an infection event with probability $\lambda/(\lambda+\tau)$, and a healing event otherwise.

For a healing event, the update depends on whether the selected site belongs to a pocket. If it is inside a pocket (i.e., if it has $D=\infty$), the number of active sites in that pocket is reduced by one. Otherwise, we denote by $n_\ell$ and $n_r$ the number of active sites in the left and right pockets adjacent to the selected $D=0$ site. If $n_\ell+n_r>0$, one active site is removed from the left pocket with probability $n_\ell/(n_\ell+n_r)$, or from the right pocket otherwise. If $n_\ell+n_r=0$, the selected active site simply becomes inactive.

For an infection event, if the selected site is inside a pocket, the number of active sites in that pocket increases by one with probability $(n-k)/n$, where $n$ is the pocket size and $k$ the number of active sites before the event. If the selected site is not in a pocket, one of its neighbors is chosen at random. If that neighbor is a $D=0$ site, it is infected; if it belongs to a pocket, the number of active sites in the pocket increases by one with probability $(n-k)/n$. An exception occurs when $k=0$ and the opposite $D=0$ neighbor of the pocket is inactive: in this case, that neighbor becomes active while the pocket remains empty ($k$ stays equal to $0$).

\subsection{Results}

\begin{figure}
    \centering
    \includegraphics[width=\linewidth]{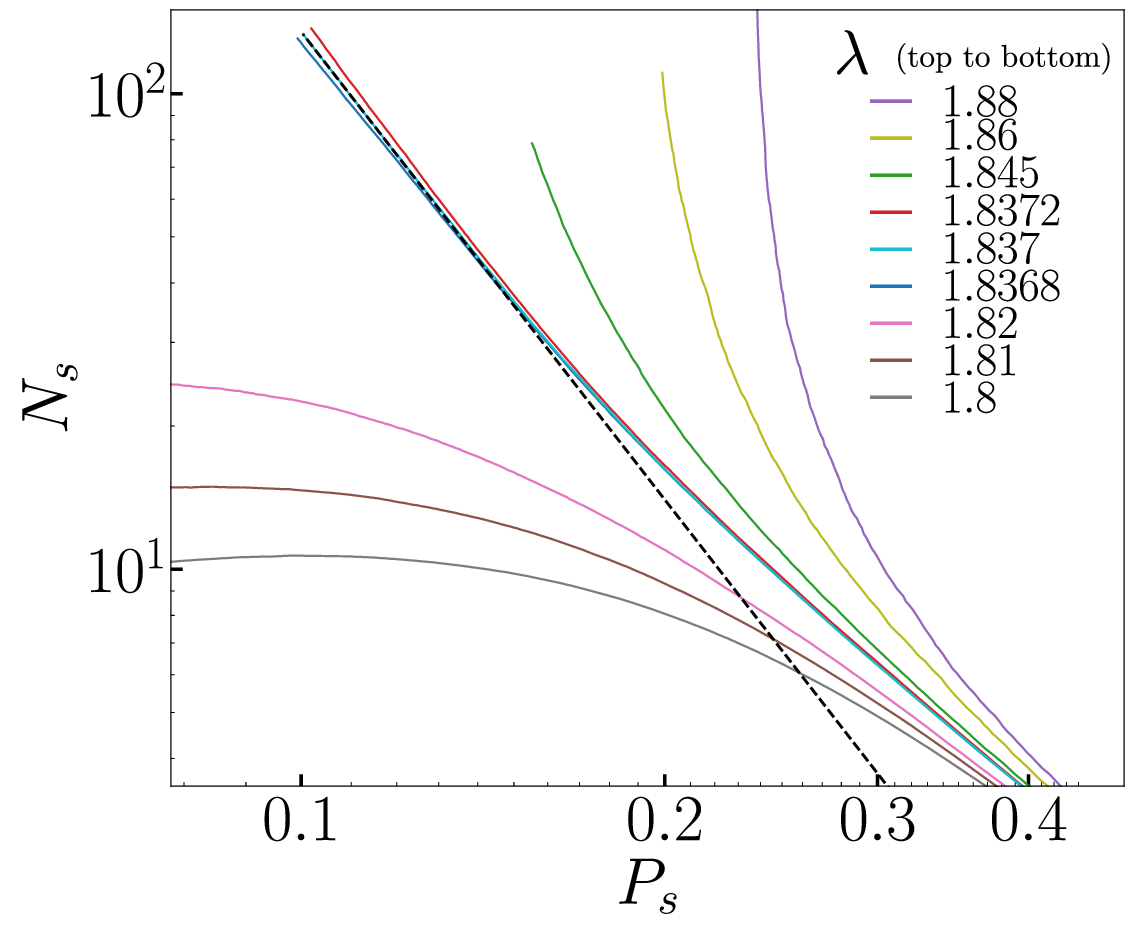}
    \caption{Number of active sites $N_{s}$ vs. survival probability $P_{s}$ for several infection rates $\lambda$ in the effective model with infinite diffusion rate. The critical curve $\lambda_{c}=1.837$ is fitted with Eq.\eqref{eq:activated_Ns_Ps} with a floating exponent,  yielding $\bar{\Theta}/\bar{\delta}=3.26(7)$. Other parameters: \(p=0.25\) and \(\tau=1\).  }
   \label{fig:Dinf_Ninfected_vs_Psur}
\end{figure}

We follow the same procedure as described in Sec.~\ref{sec:MCresults}. All simulations were performed with a uniform healing rate $\tau = 1$ and a fraction $p = 0.25$ of sites with zero local diffusion rate, using the contamination rate $\lambda$ as the control parameter. For each parameter set, at least $4\times 10^4$ disorder realizations were considered, with $30$ seed simulations per realization. The longest runs reached $t = 3\times 10^7$ with a system size $L = 10^6$, which was sufficient to avoid finite-size effects. The position of the critical point is estimated in Fig.~\ref{fig:Dinf_Ninfected_vs_Psur} by analyzing the dependence of $N_s$ on $P_s$ on a log--log scale. For both $\lambda=1.837$ and $\lambda=1.8368$, the data are approximately linear at long times. In the following, we adopt $\lambda=1.837$ as our estimate of the critical point, since the corresponding data are more consistent with the scaling expected at an IRFP.

\begin{figure}
    \centering
    \includegraphics[width=\linewidth]{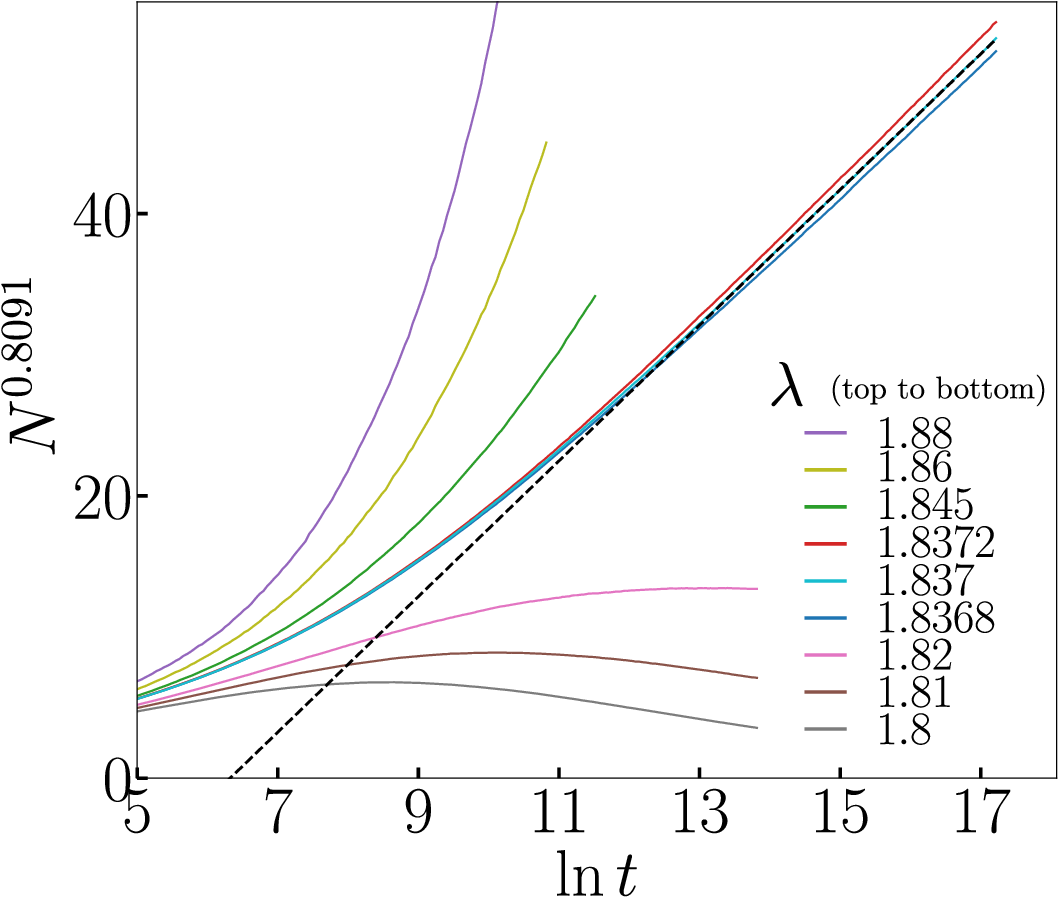}
    \caption{Number of active sites $N_{s}^{1/\bar\Theta}$ vs. $\ln t$  for several infection rates $\lambda$. The critical curve $\lambda_{c}=1.837$ is fitted with Eq.\eqref{eq:asymptotic_R_Ns} with the fixed exponent ${1/\bar{\Theta}}\approx0.8091$. }
   \label{fig:Dinf_Ninfected_vs_t}
\end{figure}

\begin{figure}
    \centering
    \includegraphics[width=\linewidth]{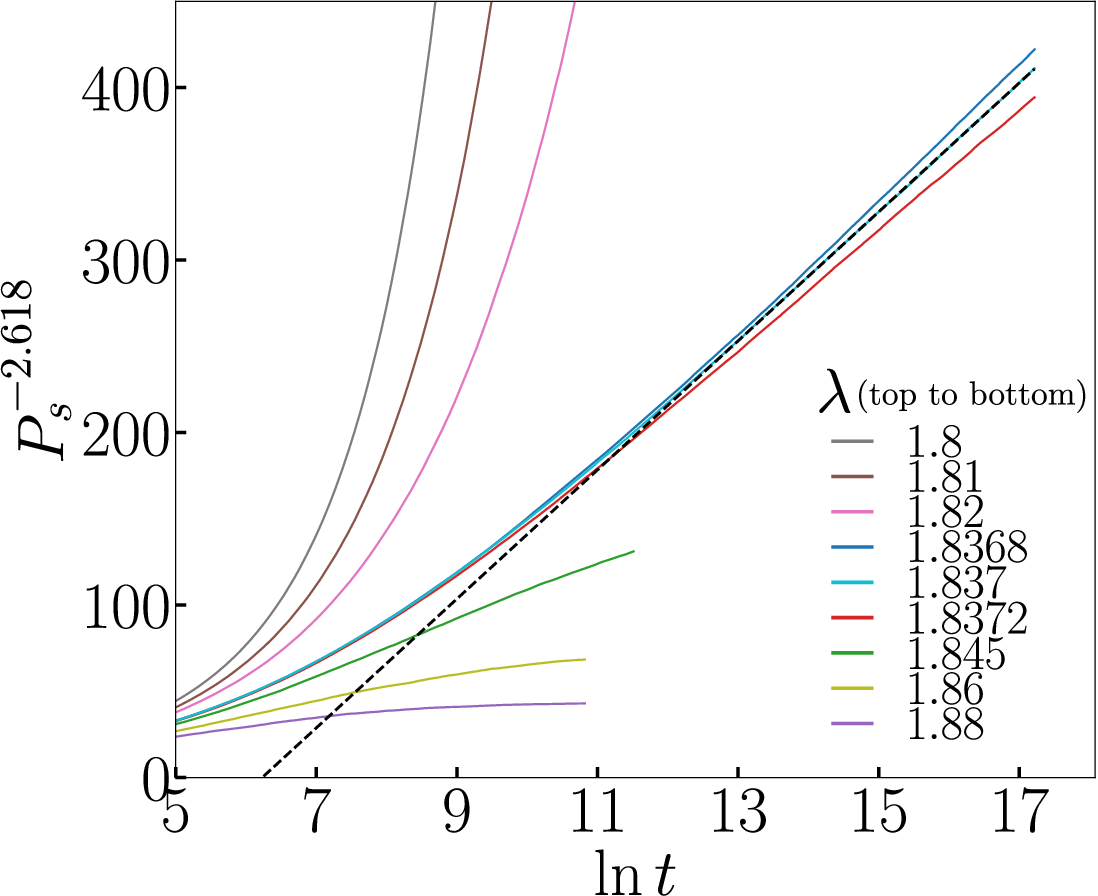}
    \caption{Survival probability $P_{s}^{-1/\bar\delta}$ vs. $\ln t$  for several infection rates $\lambda$. The critical curve $\lambda_{c}=1.837$ is fitted with Eq.\eqref{eq:asymptotic_rho_Ps} with the fixed exponent ${1/\bar{\delta}}\approx2.618$ following.}
   \label{fig:Dinf_Psur_vs_t}
\end{figure}

By fitting the data with Eq.~\eqref{eq:activated_Ns_Ps}, we obtain the exponent ratio $\bar\Theta/\bar\delta = 3.26(7)$, which remains in agreement with SDRG predictions~\cite{fisher_critical_1995, hooyberghs_strong_2003,hooyberghs_absorbing_2004}. The quantities $N_s^{0.8091}$ and $P_s^{-2.618}$ as functions of $\ln t$ are shown in Figs.~\ref{fig:Dinf_Ninfected_vs_t} and \ref{fig:Dinf_Psur_vs_t}, respectively.
Fits of the critical curves at the estimated $\lambda_c$ are consistent with the data over approximately two decades in time and yield microscopic timescale estimates $\ln t_0 \approx 6.32$ and $6.23$. These values are significantly smaller than those obtained for finite $D$ in Sec.~\ref{sec:MCresults}, as expected when starting from a stronger disorder distribution.

In summary, we have developed an effective model with a specific form of diffusion disorder that is expected to generate effective random healing rates and hence to belong to the same universality class as the CP with random-mass disorder. This prediction is numerically confirmed by Monte Carlo simulations.

%%%%%%%%%%%%%%%%%%%%%%%%%%%%%%%%%%%%%%%%%%%%%%%%%%%%%%%%%%%%%%%%%%%
\section{Feynman Diagram}
\label{sec:Feynman}

To rationalize the numerically observed relevance of quenched random diffusion disorder, we return to the DP field theory and go beyond simple power counting by examining how such disorder affects the renormalization of the vertex functions. To this end, we employ standard field-theoretic perturbation theory in Fourier space. For the present discussion, we directly use the Feynman diagram representation; a detailed derivation can be found in Refs.~\cite{janssen_renormalized_1997, hinrichsen_nonequilibrium_2000, tauber_critical_2014}. 
Our convention is that time flows from right to left, incoming arrows at a vertex represent $\phi$, and outgoing arrows represent $\tilde{\phi}$. 

The diffusion-disorder vertex originates from the Fourier transform of $S_{\mathrm{dis},D}$ in Eq.~\ref{eq:Sdis_D}, which reads
\begin{equation}
\begin{split}
    \hat{S}_{\mathrm{dis},D} 
    = \frac{\sigma_D}{2} \int &d^dk_1 \cdots d^dk_4 
       \int d\omega_1 \cdots d\omega_4 \;
       k_1^2 k_3^2 \\
       &\times \tilde{\phi}(k_2,\omega_2)\phi(k_1,\omega_1)
       \tilde{\phi}(k_4,\omega_4)\phi(k_3,\omega_3)
    \\
    &\times 
       \delta(\omega_1+\omega_2)\,
       \delta(\omega_3+\omega_4)\,
    \\
    &\times 
       \delta(k_1+k_2+k_3+k_4).
\end{split}
\label{eq:FourierSdis}
\end{equation}

\begin{figure}[t]
    \centering
    \includegraphics[width=0.9\linewidth]{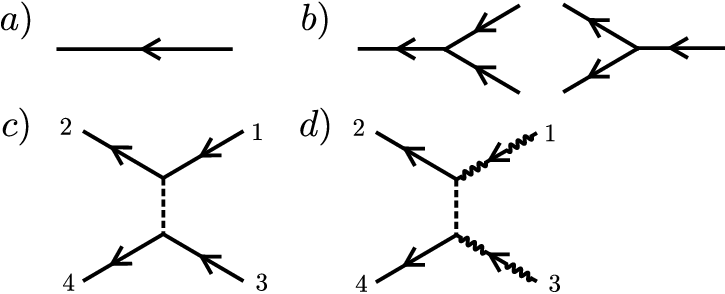}
    \caption{a) Propagator, b) interaction vertices, c) mass-disorder vertex and d) diffusion-disorder vertex used in the perturbation expansion. The factors of $k_i^{2}$ appearing in Eq.~\ref{eq:FourierSdis} are represented by wavy external legs, which makes it possible to visually distinguish the diffusion-disorder vertex from the random-mass vertex.}
    \label{fig:FGraph}
\end{figure}

The Fourier transform of $S_{\mathrm{dis},r}$ is identical but lacks the explicit  \(k_1^2\) and \(k_3^2\) factors inside the integral. To keep this distinction transparent, we mark the legs $i$ that carry a $k_i^2$ dependence with a wavy line. 
Lastly, the distinct structure of the momentum and frequency constraints is indicated by a dashed line in the graphical representation of these vertices. The propagator, interaction vertices, the mass-disorder vertex and the diffusion-disorder vertex are shown in Fig.~\ref{fig:FGraph}.

\begin{figure}[t]
    \centering
    \includegraphics[width=0.9\linewidth]{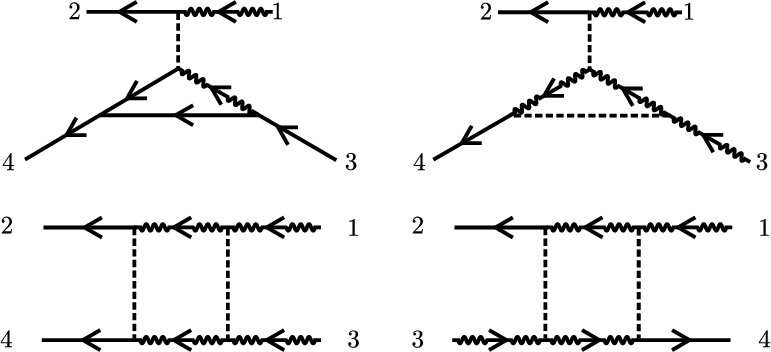}
    \caption{One-loop four-vertex diagrams, constructed by analogy with those in Ref.~\cite{janssen_renormalized_1997}. Each diagram contains at least one wavy external leg, corresponding to one factor of $k_i^2$ originating from the diffusion-disorder vertex.}
    \label{fig:oneloop}
\end{figure}

The goal here is not to perform a full RG analysis including diffusion disorder, but to show that this disorder generates random-mass disorder under renormalization. Although the topologies of the diagrams coincide with those of DP with random-mass disorder, the $k_i^2$ dependence modifies which operators are renormalized. In particular, as shown in Fig.~\ref{fig:oneloop}, all one-loop four-point diagrams involving diffusion disorder still carry residual $k_i^2$ factors, that is, at least one external leg is associated with a wavy line, and therefore do not contribute to the renormalization of the random-mass term.

\begin{figure}[t]
    \centering
    \includegraphics[width=0.4\linewidth]{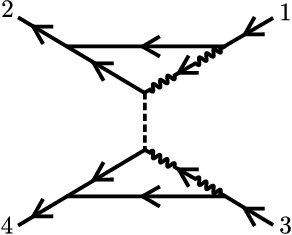}
    \caption{Example of a two-loop diagram with no wavy external legs, which therefore contributes to the renormalization of the vertex function associated with the random-mass term.}
    \label{fig:twoloop}
\end{figure}

However, at two-loop order one obtains diagrams in which the factors of $k_i^2$ are integrated out to produce a momentum-independent contribution, thereby renormalizing the vertex associated with the random-mass disorder. An example is shown in Fig.~\ref{fig:twoloop}.

Irrespective of the numerical prefactor, the appearance of such a renormalized random-mass term leads to a runaway RG flow and thus precludes the existence of a stable fixed point in the perturbative RG analysis~\cite{janssen_renormalized_1997}. Consequently, although diffusion disorder is irrelevant at the DP fixed point by power counting, it can nevertheless alter the critical behavior because, under renormalization, it generates the vertex function associated with the random-mass term. 

%%%%%%%%%%%%%%%%%%%%%%%%%%%%%%%%%%%%%%%%%%%%%%%%%%%%%%%%%%%%
\section{Conclusion}

To summarize, we have investigated the effects of quenched spatial diffusion disorder on the nonequilibrium phase transition in the CP. A simple power-counting analysis of the corresponding field theory predicts that quenched spatial diffusion disorder is irrelevant. More generally, the same conclusion holds for a broad class of reaction--diffusion processes. Nevertheless, our numerical simulations of the CP with such disorder demonstrate (i) that diffusion disorder is a relevant perturbation of the clean DP critical point, and (ii) that the transition in the presence of diffusion disorder is governed by an IRFP in the same universality class as the CP in the presence of disorder in the infection or healing rates. Specifically, our simulation data agree well with the predictions of the SDRG theory~\cite{fisher_critical_1995,hooyberghs_strong_2003,hooyberghs_absorbing_2004}.

To rationalize these findings, we developed an effective model with a binary diffusion disorder, taking either vanishing or infinite diffusion rates, and showed that its behavior closely resembles that of a contact process with random healing rates. Numerical simulations of this effective model confirm that its critical properties are indeed the same as those obtained for random healing rates.

Finally, we demonstrate that the same mechanism also arises within the field-theoretic description. We identify diagrams corresponding to an effective random-mass term generated through the renormalization by the random-diffusion term. Hence, despite being irrelevant according to power counting, diffusion disorder can generate, under renormalization, a random-mass disorder, which is known to be relevant~\cite{janssen_renormalized_1997}.

The mechanism of a formally irrelevant operator creating relevant terms under renormalization can also be observed in other problems.
For example, in a simple $\phi^4$ theory, disorder in the quartic term is irrelevant by power counting (in three dimensions), but it generates relevant random-mass disorder under renormalization. Analogously, Schwarz et al. \cite{schwartz_vycor_1993} showed that highly relevant uncorrelated random fields are generated under renormalization from less relevant anticorrelated random fields. 

Our results are also compatible with the nonperturbative version of the Harris criterion \cite{CCSF_Harris_1986}. It states that the correlation length exponent must obey the inequality $d\nu_\perp>2$ whenever tuning the disorder modifies the location of the critical point. This is fulfilled in our problem because the critical infection rate changes significantly with the concentration $p$ of sites with zero diffusion, as was verified at the beginning of Sec.\  \ref{sec:MCresults}. Consequently, the clean DP critical behavior, which violates this inequality, must change.

The relevance of random diffusion disorder raises two central questions: How can its relevance be predicted, and can it lead to critical properties that differ from those associated with random-mass disorder? For the CP, we have shown that the relevance of diffusion disorder is directly tied to that of random-mass disorder which, in turn, is predicted by the Harris criterion~\cite{harris_effect_1974}, and that the resulting critical behavior is identical to that of the random-mass case. Whether this conclusion holds more generally, however, remains an open problem. There exist many nonequilibrium models in which diffusion plays a nontrivial or even dominant role. For instance, the critical properties of the pair-contact process with diffusion remain debated, with some results suggesting a dependence on the diffusion rate~\cite{henkel_non-equilibrium_2004}. In the triplet annihilation process, sufficiently large diffusion can disperse clusters before annihilation occurs, thereby preventing the system from reaching the absorbing state~\cite{dickman_universality_1989}. In the DEP, the diffusion rates themselves determine the universality class of the transition~\cite{van_wijland_wilson_1998}. In such systems, introducing quenched disorder in the diffusion rates may therefore lead to behaviors that cannot be captured by random-mass disorder alone. Indeed, a recent numerical study of disordered DEP reports a suppression of the active phase or changes in the critical properties as a function of disorder strength for diffusion disorder, effects that are absent in the presence of mass disorder~\cite{anfray_relevance_2025}. 

These observations suggest that random diffusion disorder may constitute a genuinely distinct perturbation in nonequilibrium critical phenomena, calling for further analytical and numerical investigations to delineate its role and possible differences from random-mass disorder.

\section{Acknowledgments}
The Monte Carlo simulations were performed on the Pegasus and Mill \cite{mill_2024} clusters at Missouri University of Science and Technology. This work was partially supported by grants from the National Science and Technology Council, Taiwan (Grant No. NSTC 111-2112-M-001-027-MY3 and 114-2112-M-001-062) and Academia Sinica Career Development Award (Project No. AS-CDA-114-M02). VA acknowledges support from Academia Sinica Postdoctoral Scholar Program.

\appendix

\section{Survival probability of a site neighbor to a pocket of size $n=1$}
\label{app:S_proba}

We consider a chain with three sites where the two external sites are active and have hopping rate $D=0$. The middle site has $D=\infty$, i.e., it belongs to a pocket of size $n=1$.

Let $p(101,t)$ denote the probability that the configuration at time $t$ is $(1,0,1)$, and $p(111,t)$ the probability that the configuration is $(1,1,1)$. For $\alpha=2$ we consider the survival of either external site; for $\alpha=1$ that of a specific one.

The evolution of these two states (see Fig.~\ref{fig:illustration_DinfCP} for the dynamical rules) is governed by
\begin{align}
    \partial_t p(101,t) &= -(\alpha\tau+\lambda)\,p(101,t) + 3\tau\, p(111,t), \label{eq:p101-eq}\\
    \partial_t p(111,t) &= \lambda\, p(101,t) - 3\tau\, p(111,t), \label{eq:p111-eq}
\end{align}
with initial condition \(p(101,0)=1,\) and \(p(111,0)=0\). The survival probability is
\begin{equation*}
    S(t)=p(101,t)+p(111,t),
\end{equation*}
from which one immediately obtains
\begin{equation*}
    \partial_t S(t) = -\alpha\tau\, p(101,t).
\end{equation*}

Equation~\eqref{eq:p101-eq} allows us to express $p(111,t)$ in terms of $p(101,t)$ and its derivative:
\begin{equation}
    p(111,t)=\frac{1}{3\tau}\!\left(\partial_t p(101,t) + (\alpha\tau+\lambda)p(101,t)\right).
    \label{eq:p111-from-p101}
\end{equation}

Substituting Eq.~\eqref{eq:p111-from-p101} into Eq.~\eqref{eq:p111-eq} yields the second-order differential equation
\begin{equation}
\begin{split}
    \frac{\partial^2}{\partial t^2} p(101,t)
    + \big((3+\alpha)\tau + \lambda\big)\,\partial_t p(101,t)
    \\+ 3\alpha\tau^2\, p(101,t) = 0.
    \label{eq:second-order}
\end{split}
\end{equation}

The general solution of Eq.~\eqref{eq:second-order} is
\begin{equation*}
    p(101,t) = A e^{r_+ t} + B e^{r_- t},
\end{equation*}
where
\begin{align*}
    r_\pm &= -\frac{\lambda + (3+\alpha)\tau}{2} \pm \frac{\Delta}{2}, \\
    \Delta &= \sqrt{\lambda^2 + 2(3+\alpha)\tau\lambda + \tau^2(\alpha-3)^2}.
\end{align*}

The initial conditions give
\begin{align*}
    A + B &= 1,\\
    r_+ A + r_- B &= -(\alpha\tau+\lambda),
\end{align*}
so that
\begin{equation*}
    A = \frac{1}{2}\!\left(1 + \frac{(3-\alpha)\tau - \lambda}{\Delta}\right), \qquad
    B = \frac{1}{2}\!\left(1 - \frac{(3-\alpha)\tau - \lambda}{\Delta}\right).
\end{equation*}

Thus
\begin{equation}
\begin{split}
    p(101,t) =& \frac{1}{2}\!\left(1+\frac{(3-\alpha)\tau-\lambda}{\Delta}\right)e^{r_+t} \\
    &+ \frac{1}{2}\!\left(1-\frac{(3-\alpha)\tau-\lambda}{\Delta}\right)e^{r_-t}.
    \label{eq:p101-solution}
\end{split}
\end{equation}

Since $\partial_t S(t)=-\alpha\tau\,p(101,t)$, integrating using \eqref{eq:p101-solution} gives
\begin{align*}
    S(t) =& -\frac{\alpha\tau}{2r_+}\!\left(1+\frac{(3-\alpha)\tau-\lambda}{\Delta}\right)e^{r_+ t}\\
       &-\frac{\alpha\tau}{2r_-}\!\left(1-\frac{(3-\alpha)\tau-\lambda}{\Delta}\right)e^{r_- t},
\end{align*}
% Using the identities
% \begin{align*}
%     -\frac{\alpha\tau}{2r_+}\!\left(1+\frac{(3-\alpha)\tau-\lambda}{\Delta}\right)
%     -\frac{\alpha\tau}{2r_-}\!\left(1-\frac{(3-\alpha)\tau-\lambda}{\Delta}\right) &= 1,\\
%     -\frac{\alpha\tau}{2r_+}\!\left(1+\frac{(3-\alpha)\tau-\lambda}{\Delta}\right)
%     +\frac{\alpha\tau}{2r_-}\!\left(1-\frac{(3-\alpha)\tau-\lambda}{\Delta}\right)
%     &= \frac{\lambda + (3-\alpha)\tau}{\Delta},
% \end{align*}
which simplifies to the compact form
\begin{equation}
\begin{split}
    S(t)= e^{-\frac{\lambda + (3+\alpha)\tau}{2}t} \Bigg[ &\cosh\!\left(\frac{\Delta t}{2}\right)\\
    &+ \frac{\lambda + (3-\alpha)\tau}{\Delta}\, \sinh\!\left(\frac{\Delta t}{2}\right) \Bigg].
    \label{eq:S-final} 
\end{split}
\end{equation}

\begin{figure}
    \centering
    \includegraphics[width=\linewidth]{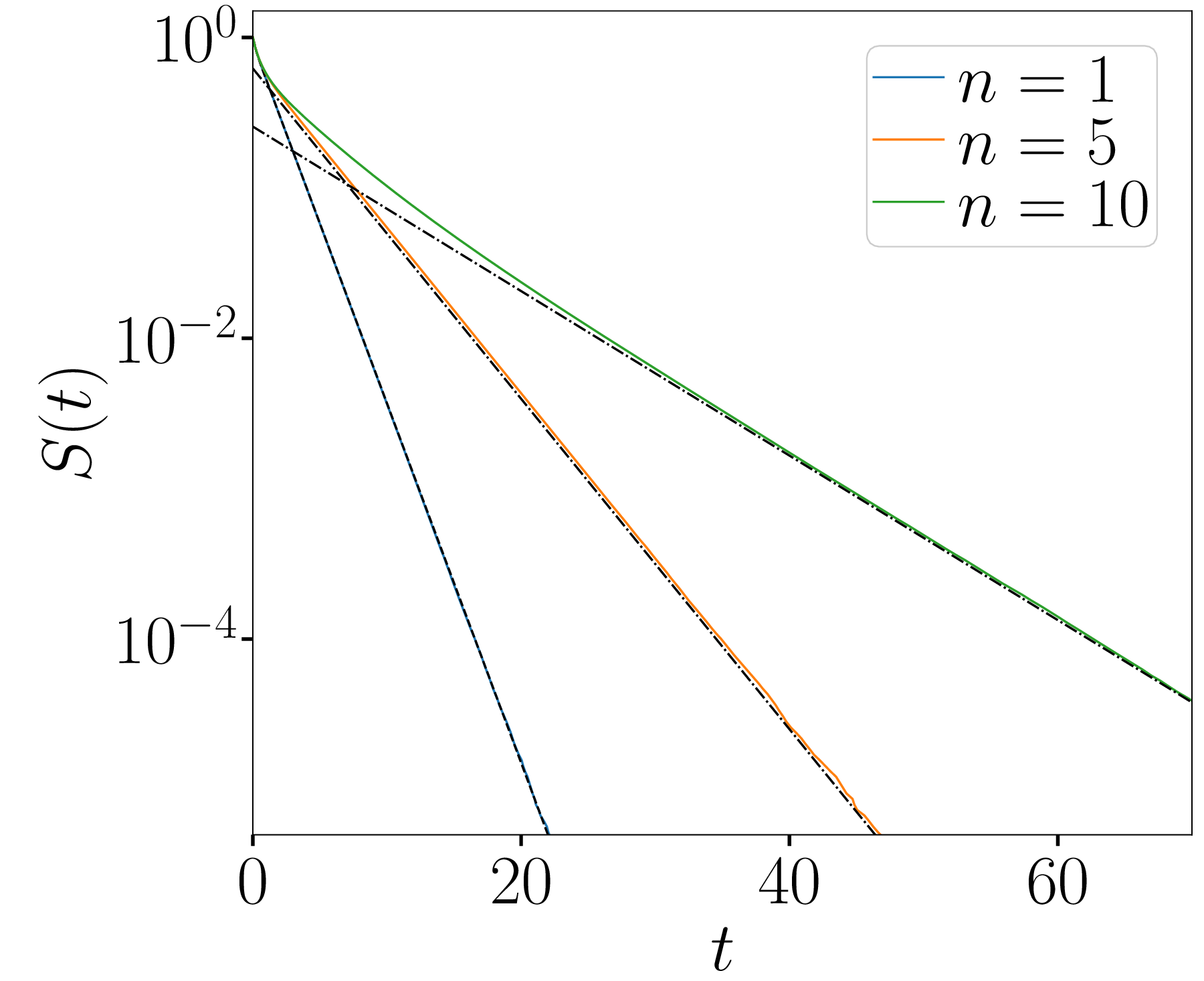}
    \caption{Time evolution of the survival probability $S(t)$ of an active site with $D=0$ adjacent to a pocket of size $n$ with initially no active site (the other external site is assumed to always stay active, corresponding to $\alpha=1$), with $\lambda=2$ and $\tau=1$. The probability $S(t)$ is obtained from a Gillespie simulation of the dynamical rules shown in Fig.~\ref{fig:illustration_DinfCP}b). The black dotted line corresponds to Eq.~\eqref{eq:S-final}, while the dash-dotted lines show the exponential decay with rate $\tilde{\tau}$ obtained from the smallest eigenvalue of the associated Liouville operator (see Appendix~\ref{app:LiouvilleDecay}).}
    \label{fig:SurvivalProbabilityOneNeighbor}
\end{figure}

To verify this result, we show in Fig.~\ref{fig:SurvivalProbabilityOneNeighbor} the survival probability $S(t)$ of a single active external site in the presence of one active neighbor for several values of $n$. The theoretical prediction is plotted in black. For $n=1$, Eq.~\eqref{eq:S-final} fits perfectly, whereas for $n>1$ we plot the late-time exponential decay given by the largest eigenvalue of the Liouville operator (see Appendix~\ref{app:LiouvilleDecay}). 

\section{Late-time decay of the survival probability}
\label{app:LiouvilleDecay}
To determine the late-time decay of the survival probability of an active site adjacent to a pocket of size $n$ (assuming the other external site is also active), we study the time evolution of the probability vector
\begin{equation}
    \partial_t \ket{P_t} = -\mathcal{L}\,\ket{P_t},
    \label{eq:Liouville}
\end{equation}
where $\mathcal{L}$ is the Liouville operator constructed from the dynamical rules described in Fig.~\ref{fig:illustration_DinfCP}. Each component of $\ket{P_t}$ corresponds to the probability of a specific configuration of the pocket and the external sites.

As an explicit example, consider a pocket of size $n=2$. The accessible configurations form a three-state system, and the corresponding Liouville operator is
\begin{equation*}
    \mathcal{L} = -
        \begin{bmatrix}
        -(\lambda+\alpha\tau) & 3\tau & 0 \\
        \lambda & -(3\tau+\lambda) & 4\tau \\
        0 & \lambda & -4\tau
        \end{bmatrix},
\end{equation*}
where $\alpha=1$ or $2$ depending on whether we consider the survival of one specific external site ($\alpha=1$) or of either of the two indistinguishably ($\alpha=2$).

Because the sum of the elements in the first column is not zero, probability is not conserved: the system eventually reaches the absorbing configuration where the sum of probabilities is zero. Consequently, the smallest eigenvalue of $\mathcal{L}$ is strictly positive. Denoting this lowest eigenvalue by $E_0$, the survival probability satisfies $S(t) \sim e^{-E_0 t}$ as $t\to\infty$ which defines the effective decay rate as $\tilde{\tau} = E_0$.

For any finite pocket size $n$, the dimension of $\mathcal{L}$ is finite and therefore the late-time decay is always exponential. One may wonder whether an infinite pocket could give rise to a non-exponential decay. However, pockets have sizes drawn from a geometric distribution with $0<p<1$, and are therefore finite. As a result, the survival probability always decays exponentially at sufficiently late times.

\bibliography{references_tv}
\end{document}